\documentstyle[graphicx]{elsart}
\begin{document}
\begin{frontmatter}
\title{Symmetry consideration and $e_g$ bands \\ in NdNiO3 and YNiO3}
\author{Susumu Yamamoto and Takeo Fujiwara}
\address{Department of Applied Physics,  University of Tokyo,
Tokyo 113-8656, Japan}
\begin{abstract}
Group theoretical analyses are applied to the magnetic and electronic 
structures of NdNiO$_3$ and YNiO$_3$, 
whose electronic structures have been studied very recently by the
LSDA+U method. 
The space groups of the crystallographic structure of these
materials are $Pbnm$ and $P2_1/n$, respectively.\cite{RNiO3NDIFF,YNiO3}
There are no Jahn-Teller distortion modes consistent with observed
propagation vector of spin polarized neutron-diffraction. 
Two possible spin configurations (magnetic space groups)  
$C_amc2_1$ and $P_ba$ are derived from
$Pbnm$ for NdNiO$_3$. 
Then the electronic structures with each symmetry are studied
by using the tight-binding model. 
Clear separation of partial density of states
exists in $P_ba$ but does not in orthorhombic $C_amc2_1$.
The combination of results of the present work and the previous LSDA+U
calculation shows that the ground state magnetic structure 
of NdNiO$_3$ is monoclinic $P_ba$.  
NdNiO$_3$ has only one kind of Ni sites, while YNiO$_3$ has
two inequivalent Ni sites. The magnetic structure of YNiO$_3$ is
monoclinic $P_b2_1/a$, if one assumes no magnetic moment on
Ni ions (Ni$^{4+}$) on one of these inequivalent sites.
\end{abstract}

\end{frontmatter}
The perovskite nickelates NdNiO$_3$ and YNiO$_3$ 
are anti-ferromagnetic insulators in their low temperature phases. 
Their magnetic structures are  characterized by a propagation 
vector $\vec{k}=(\frac{1}{2},0,\frac{1}{2})$~\cite{RNiO3MagDIFF} 
and, therefore, the magnetic unit cell  is 
a $2 \times 1 \times 2$ supercell of the original 
crystallographic unit cell.
The spin alignment in these two nickelates is quite unique one
such that the symmetrical coexistence of ferro- and antiferromagnetic 
couplings  along the three pseudo cubic axes as   $\rm
\cdots Ni^\uparrow -\stackrel{\mbox{\tiny FM}}{O}-Ni^\uparrow -
\stackrel{\mbox{\tiny AFM}}{O}-Ni^\downarrow -\stackrel{\mbox{\tiny
FM}}{O}-Ni^\downarrow\cdots$. 
In YNiO$_3$, the charge disproportionation 
2Ni$^{3+}\rightarrow $ Ni$^{2+}+$Ni$^{4+}$ and 
the symmetrical coexistence of ferro- and antiferromagnetic 
couplings are both observed, though the Ni$^{4+}$ ions have small 
magnetic moments.~\cite{YNiO3}
Diffraction experiments in NdNiO$_3$ do not show  
accompanied Jahn-Teller distortion associated with the orbital
polarizations of Ni $e_g$ orbitals.

Very recently, we have calculated the electronic structures of 
the perovskite nickelates NdNiO$_3$ and YNiO$_3$ 
by the LSDA+U method which will be published elsewhere.~\cite{RNiO3LSDAU} 
We observed in that study no orbital ordering in NdNiO$_3$ and 
the charge disproportionation in YNiO$_3$. 
The basic consideration by the group theory about the magnetic
structures, the possibility of the Jahn-Teller modes and the simple
picture of the $e_g$ bands in these structure 
are given   in the present report. 


{\bf \it Crystallographic  space group}:
The crystallographic space groups of NdNiO$_3$ and YNiO$_3$ are
orthorhombic $Pbnm$ and monoclinic $P2_1/n$, respectively.
$P2_1/n$ is the subgroup of $Pbnm$ and their group elements are as
follows; 
$Pbnm =  P2_1/n \otimes \{E, \{\sigma_x|\frac{1}{2}\frac{1}{2}0\}\}$,
$P2_1/n = \{E,\{\sigma_y|\frac{1}{2}\frac{1}{2}\frac{1}{2}\}, I,
  \{ C_{2y}|\frac{1}{2}\frac{1}{2}\frac{1}{2} \}\} $.
where $E$ is the identity operator,  $I$ the  inversion operator, 
$\sigma_x$ the mirror operation with respect to the $y$-$z$ plane and 
$C_{2y}$ the two-fold rotation  operator  with respect to the $y$ axis. 
Both  NdNiO$_3$ and YNiO$_3$ systems contain four Ni  
ions in their crystallographic unit cells and each locates on 
the inversion center. 
Four Ni sites are  labeled as 
Ni(1)=$(\frac{1}{2},0,0)$, 
Ni(2)=$(0,\frac{1}{2},0)$,
Ni(3)=$(\frac{1}{2},0,\frac{1}{2})$ and 
Ni(4)=$(0,\frac{1}{2},\frac{1}{2})$.
In NdNiO$_3$, all four nickel positions are equivalent.
In YNiO$_3$, no symmetry operations 
transfer a position Ni(1) or Ni(4)  to that of Ni(2) or Ni(3). 
Therefore, there  are two inequivalent positions of Ni ions 
in YNiO$_3$.


{\bf \it Magnetic space group}:
First we consider the magnetic space groups generated
from the crystalline space group $Pbnm$ with
the experimentally observed propagation vector
$\vec{k}=(\frac{1}{2},0,\frac{1}{2})$.
The translation vector along the $\alpha$-axis of crystallographic $Pbnm$
structure is denoted by  $\vec{T_\alpha}$. 
The system is the orthorhombic base centered lattice and 
the (orthogonal)  primitive translation vectors should be 
$\{2\vec{T_x},\ \vec{\vec{T_y}},\ 2\vec{T_z}\}$. 
The nonorthogonal translation vectors 
$\vec{T_x}+\vec{T_z}$ and $ \vec{T_x}-\vec{T_z}$ 
transform lattice points at corners of the Bravais lattice 
to those at the base centers. 


Assuming the orthorhombic magnetic unit cell, 
one of possible spin arrangements is the structure
where Ni(1), Ni(2), Ni(3), Ni(4) in a unit cell of $Pbnm$ 
have the same spins
and the translation operators $\vec{T_x}$ or $\vec{T_z}$ flips
the spins of Ni ions.~\cite{RNiO3MagDIFF}
Then the ferromagnetically aligned spins are confined within a one-dimensional rod 
extending towards $y$-direction.
The doubled translation vectors along the $x$- and $z$-axes 
prohibit the screw operator $2_1$ along the $x$- and $z$-axes.~\cite{CaV2O4} 
Then the unitary part of the resultant magnetic space group should be
$Pmc2_1= \{E, \{C_{2y}|\frac{1}{2}\frac{1}{2}\frac{1}{2}\},
\{\sigma_x|\frac{1}{2}\frac{1}{2}0\}, \{\sigma_z|00\frac{1}{2}\} \}$. 
An augmenting operator with the time-reversal operator is the translation
one $\vec{T_x}$ and the magnetic translation vector should be
$2\vec{T_x}$. 
The resultant magnetic space group is base centered orthorhombic $C_amc2_1$.

Another possible spin arrangement associated with $Pbnm$ and 
$\vec{k}=(\frac{1}{2},0,\frac{1}{2})$
is that where the same spins are located on Ni ions 
in a parallelepipedon spanned by translation operators
$\{\vec{T_x}, \vec{T_y}, \vec{T_z}-\vec{T_x}\}$.
In this structure, the two-fold rotational operator with respect to
the $z$-axis and the mirror operator with respect to the $x$-$y$ plane
are prohibited because these contradict to the monoclinic Bravais
lattice. 
The screw operator $2_1$  along $x$-axis is also prohibited due
to cell doubling. 
There remain only two symmetry operators $E$ and
$\{\sigma_y | \frac{1}{2}\frac{1}{2}\frac{1}{2}\}$ the
glide operator along the monoclinic axis $\vec{T_z}-\vec{T_x}$. 
Because the augmenting operator with the
time-reversal operator is the translation $\vec{T_x}$, the resultant
magnetic space group is monoclinic $P_ba$. 
The magnetic structure of NdNiO$_3$ with $P_ba$ symmetry
is of double layers of the (101) plane, on which all Ni ions have
the same spins.

In YNiO$_3$,
 there are two inequivalent Ni sites,
as already stated.
If there exists no magnetic moment on Ni sites, e.g. Ni(2) and Ni(3), 
the system  has the inversion symmetry around the sites of
Ni(1) and Ni(4), and the magnetic space group should be $P_b2_1/a$.
This is actually the case for YNiO$_3$ where the charge disproportionation 
2Ni$^{3+}\rightarrow $ Ni$^{2+}+$Ni$^{4+}$ occurs and the magnetic
moment of Ni$^{4+}$ is zero. 


{\bf \it Jahn-Teller distortion}: 
In many transition metal perovskites, the Jahn-Teller distortion 
coexists with the complicated spin and orbital orderings. 
In perovskite nickelates, no Jahn-Teller distortion has been 
reported yet.
A possibility of  the Jahn-Teller mode consistent with 
the space group $Pbnm$ and the propagation vector 
$\vec{k}=(\frac{1}{2},0,\frac{1}{2})$ can be discarded in a following 
way.
A unit cell of $Pbnm$ contains four NiO$_6$ octahedra which share 
O ions with each other.  
Let us consider the $E_g$ Jahn-Teller modes in two octahedra 
Ni(1)O$_6$ and  Ni(4)O$_6$ which do  share  no O ions.
There are two two-dimensional irreducible representations in the 
${\vec k}$-group $G({\vec k}=(\frac{1}{2},0,\frac{1}{2}))$,
and they are listed in Ref. \cite{RNiO3MagDIFF}.
The only elements of non-zero character are 
$E$ and $\{\sigma_y|\frac{1}{2}\frac{1}{2}\frac{1}{2}\}\}$ 
and $\chi(\{\sigma_y|\frac{1}{2}\frac{1}{2}\frac{1}{2}\}\})=\pm 2$. 
Therefore, the projection operator on those representations are
$P=d/g\sum_{R\in G}\chi(R)^*R= \frac{1}{2}
(\{E|000\}\}\pm\{\sigma_y|\frac{1}{2}\frac{1}{2}\frac{1}{2}\}\}$. 
When 
$\{\sigma_y|\frac{1}{2}\frac{1}{2}\frac{1}{2}\}\}$ operates on $Q^1u$,
resultant mode equals to $Q^4u$ and vice versa.
Here $Q^ju$ is the $E_g$ $u$ mode of Ni($j$)O$_6$ ($j=1,\ 4$). 
Then basis vectors of the lattice distortion 
that belong to  these irreducible representations are $Q^1u\pm Q^4u$ 
in  two octahedra Ni(1)O$_6$ and  Ni(4)O$_6$.
The propagation vector $\vec{k}=(\frac{1}{2},0,\frac{1}{2})$ 
requires  that the translation operation of  
$\vec{T_x}$ or $\vec{T_z}$ changes the sign of a displacement vector. 
The resultant displacement vectors are shown in Fig.~\ref{FigMode}.  
Now the displacements of all O ions of Ni(3)O$_6$ have 
been determined, because they are contained either in Ni(1)O$_6$ 
or in   Ni(4)O$_6$. 
The resultant distortion of the Ni(3)O$_6$ octahedron is 
an ungerade mode, as seen in Fig.~\ref{FigMode}, 
and cannot be generated from $E_g$ mode on this octahedron.
The same arguments can be applied to other distortion modes. 
This is the proof that the $E_g$ mode in one octahedron cannot be 
transferred to all octahedra over the whole unit cell.

{\bf \it Structure of $e_g$ bands}:
Because the 3d $e_g$ orbitals in Ni ions are partly filled, 
the highest occupied (HOMO) states should be  $e_g$ bands. 
Since the structures of NdNiO$_3$ and YNiO$_3$ have been discussed 
group theoretically, the band structure of the $e_g$ orbitals 
can be analyzed from the similar group theoretical view point and
this analysis would be useful to understand the calculated 
electronic structure of 
these systems by the LSDA+U.~\cite{RNiO3LSDAU}
The discussion can be based on a tight-binding Hamiltonian 
of single spin component and 
one can neglect electron transfer from one spin site to sites with 
opposite spin.  
Since there are no Jahn-Teller modes and the tilting angle  
of the NiO$_6$ octahedra is small,
distortion can be neglected  and  an ideal cubic structure is
assumed. 
The Hamiltonians do not have any terms of the Slater-Koster parameter 
$V_{dd\pi}$ due to this high symmetry.

First, let us discuss the case of $C_amc2_1$. 
Electrons can
transfer only within a one-dimensional rod and the ferromagnetic
cell extending along the $y$-axis.  The $e_g$ orbitals are located
at  four vertices of a unit cube, i.e $(0,0,0)$, $(0,0,1)$,
$(1,0,0)$, $(1,0,1)$, with a translation vector $\vec{t}=(1,1,0)$. 
The parameter $V_{dd\sigma}$ and
$V_{dd\delta}$  are chosen to equal to $-1$ and $-0.1$, respectively, and the
diagonal atomic energy to be zero. 
Figure \ref{FigBandC} shows the resultant energy bands $E(k)$
in comparison with  those 
by the LSDA+U method.~\cite{RNiO3LSDAU} Because
this one-dimensional lattice is bipartite, the $E(k)$ curves are
symmetric with respect to $E=0$. As occupation in Ni case is just one
electron per each site, the Fermi energy is at the energy zero
$E_F=0$. The result of the tight-binding Hamiltonian can reproduce
the essential feature of the $e_g$ band. 
Two widely dispersive and
two narrow bands are occupied  and  they have no dispersion
along $x$- and $z$-axis. 
Two dispersive curves are crossing at the
Fermi level. 
The $\vec{k}$-value of the crossing point actually depends on  
the ratio $V_{dd\delta}$/$V_{dd\sigma}$, 
because the two-fold degeneracy at
the Fermi level is not due to the crystallographic symmetry.
The separation of partial density of states can be seen very clearly 
in the partial density of
states in the present tight binding model, as shown in Fig.~\ref{FigBandC}c.
The model Hamiltonian keeps much higher symmetry than the actual
system and the actual  system  opens the gap as seen in Fig.~\ref{FigBandC}a

Next, let us consider the case of $P_ba$. 
The (101) plane of the  $Pbnm$ structure corresponds  
to the (111) plane of the cubic lattice. 
Hereafter, the subscript ``{\it c}'' denotes the cubic lattice as (111)$_c$.  
Centers of atomic $e_g$ orbitals (Ni ions)  locate at
vertices of a unit cube; $(0,0,0)$ and $(0,0,1)$ 
with a translation vector  $\vec{a}=(1,0,-1)$ and $\vec{b}=(0,1,-1)$. 
Since the system has the three-fold rotational symmetry around
[111]$_c$ axis, the Brillouin zone is a hexagon.
The unit cell contains two Ni atoms and, therefore, 
four bands exist in the Brillouin zone. 
Two of them have the  dispersion 
proportional to the $V_{dd\delta}$ and the
other two have larger dispersion. 
They  touch each other at the Fermi energy $E_{\rm F}$. 
Each band is symmetric around 
$E_F$, because the lattice is bipartite. 
The contour plot of the band energy $E(\vec{k})$ 
touching $E_F$ is shown in Fig.~\ref{FigBandD}, 
where the $V_{dd\sigma}$ and $V_{dd\delta}$ are set to be 
$-1.0$ and $-0.1$ respectively,  
together with the band structures  of the LSDA+U calculation.~\cite{RNiO3LSDAU} 
The Fermi surface is reduced to just points,  
six vertices of the first Brillouin zone (BZ), and
there is two-fold degeneracy at the Fermi surface
due to three-fold rotational symmetry.
Therefore, this system is a marginal metal. 
The eigen states at the vertex of the BZ 
are corresponding to bonding and anti-bonding states of
$\varphi_{u\pm}$ on each adjacent plane,
where $\varphi_{u+}=-\frac{1}{\sqrt{2}}(\varphi_{3z^2-r^2}+{\rm i}
\varphi_{x^2-y^2})$ and
$\varphi_{u-}=\frac{1}{\sqrt{2}}(\varphi_{3z^2-r^2}-{\rm i}
\varphi_{x^2-y^2})$.
This is not only the case at the Fermi energy. 
The system of the model Hamiltonian is the trigonal D$_{3d}$ 
and the basis wavefunctions are also  $\varphi_{u\pm}$, 
which are the basis of the $E_g$ representation. 
Therefore, this system does not cause the orbital polarization at all, 
and the partial density of states by the tight-binding model is shown 
in Fig.~\ref{FigBandD}c.
Because this system is a marginal metal, the small perturbation 
easily lifts the two-fold degeneracy at the Fermi surface
and leads the system to the insulating phase.
Actually, the gap positions in Fig.~\ref{FigBandD}a are identical 
to the vertices of the BZ 
in Fig.~\ref{FigBandD}b and the system has no 
three-fold symmetry.



Both simple tight-binding models based on $C_amc2_1$ and $P_ba$ spin alignment,
give metallic bands
in contradiction to the fact that NdNiO$_3$ is insulator, 
simply because the models have much higher symmetry than that of re
realistic models. 
The result of LADA+U calculation, which is referred before, shows that
the bands in $P_ba$ open the gap of 0.11~eV without 
orbital polarization and those in $C_amc2_1$ also open the gap of 0.24~eV 
with a slight orbital order.~\cite{RNiO3MagDIFF}
Since the orbital polarization has not been observed and no
Jahn-Teller distortion reported, the spin alignment of $P_ba$ is
probably the case. 

The charge disproportionation $2 \mbox{Ni}^{3+} \rightarrow
\mbox{Ni}^{2+} + \mbox{Ni}^{4+}$ is observed experimentally 
in YNiO$_3$.
This is accompanied by the crystallographic symmetry reduction from $Pbnm$ to $P2_1/n$.
If the spin  on  Ni$^{4+}$ ions is  zero, then  only one  magnetic
structure $P_b2_1/a$ is allowed. 
In this case, the  charge polarization (charge order) opens the gap. 
Our LSDA+U calculation gives the band gap of 1.03~eV in YNiO$_3$.~\cite{RNiO3LSDAU}
Experimentally observed magnetic moment on  Ni$^{4+}$ ions is not
zero and this discrepancy may be due to the dynamical effect of
spins.~ \cite{YNiO3}

{\bf \it Conclusion}: 
We discussed very details of the magnetic space group for 
NdNiO$_3$ and YNiO$_3$, 
and discarded the possibility of the Jahn-Teller distortion 
extending over the whole magnetic cell. 
The electronic structures of $e_g$ bands are analyzed by the tight-binding
Hamiltonian, and compared with the calculated results by the LSDA+U
method.~\cite{RNiO3LSDAU}

The present tight-binding model, in both cases of NdNiO$_3$ and
YNiO$_3$, is so simple and several symmetry elements
are not included in the Hamiltonian. 
Then the resultant band structure is of the marginal metal but not of
insulator.  
This discrepancy is not serious at all. 

{\bf \it Acknowledgment:}
This work is supported by
Grant-in-Aid for COE Research ``Spin-Charge-Photon''.


\begin{figure}
	\begin{center}
	\end{center}
	\caption{The displacement vectors corresponding to a basis function
	of group of $\vec{k}=(\frac{1}{2},0,\frac{1}{2})$ of $Pbnm$.
	Center of each octahedron is occupied by Ni ion;
	Ni(1)=$(\frac{1}{2},0,0)$, 
  Ni(2)=$(0,\frac{1}{2},0)$,
	Ni(3)=$(\frac{1}{2},0,\frac{1}{2})$,
	Ni(4)=$(0,\frac{1}{2},\frac{1}{2})$. 
 The Ni ions with  prime marks can be obtained from the original (without a 
 prime) ions by  the primitive translations.
    }
\label{FigMode}
\end{figure}
\begin{figure}
	\begin{center}
	\end{center}
	\caption{  Energy bands $E(\vec{k})$  for one spin conponent in
	NdNiO$_3$ with $C_amc2_1$ by the LSDA+U method  (a) and those by the tight binding
model (b). 
  The Slater-Koster parameters $V_{dd\sigma}$ and $V_{dd\delta}$
eqaual to  $-1.0$ and $-0.1$ in (b),  respectively. 
(c) The partial density of states of $e_g$ states in NdNiO$_3$ by the 
tight binding model. The orbital polarization can be clearly seen.}
\label{FigBandC}
\end{figure}
\begin{figure}
	\begin{center}
	\end{center}
	\caption{  The energy bands $E(\vec{k})$  for 
  spin strucrutre  $P_ba$ of NdNiO$_3$ by the LSDA+U method
(a) and those by the tight binding model (b). The results (b)
should be folded along chain line and the resulting E-$\vec{k}$ curves
correspond to those in (a).
  The Slater-Koster parameter $V_{dd\sigma}$ and $V_{dd\delta}$
eqaual to  $-1.0$ and $-0.1$ in (b),  respectively. 
(c) The partial density of states of $e_g$ states by the 
tight binding model. The separation of partial density of states
does not exist.
 (d) The contour plot of $E(\vec{k})$  of
	the corresponding tight-binding model Hamiltonian. 
  The $\vec{g_a}$ and
	$\vec{g_b}$ are the basis set of the  reciprocal lattice, and 
  the corresponding
	translation vectors are $\vec{a}=(1,0,-1)$ and $\vec{b}=(0,1,-1)$.
	Translation vector $\vec{c}$ and a  corresponding vector $\vec{g_c}$ are
  parallel to $(1,1,1)$. 
}
\label{FigBandD}
\end{figure}
\end{document}